# STATE PRICE DENSITY ESTIMATION VIA NONPARAMETRIC MIXTURES[1]

By Ming Yuan

*Georgia Institute of Technology*

We consider nonparametric estimation of the state price density encapsulated in option prices. Unlike usual density estimation problems, we only observe option prices and their corresponding strike prices rather than samples from the state price density. We propose to model the state price density directly with a nonparametric mixture and estimate it using least squares. We show that although the minimization is taken over an infinitely dimensional function space, the minimizer always admits a finite dimensional representation and can be computed efficiently. We also prove that the proposed estimate of the state price density function converges to the truth at a "nearly parametric" rate.

**1. Introduction.** In this paper we consider estimating the risk-neutral distribution encapsulated in option prices. Risk-neutral distributions, often characterized by state price densities, recovered from option prices reflect investors' expectation toward the future returns of the underlying assets. It manifests the preferences and risk aversion of a representative agent [Aït-Sahalia and Lo (2000); Jackwerth (2000); Rosenberg and Engle (2002)]. Consider, for example, a European call option with maturity date $T$ and strike price $X$. Under the no-arbitrage principle, its price at $t$ can be given as

$$(1.1) \qquad C(X, S_t, r_{t,\tau}, \tau) = e^{-r_{t,\tau}\tau} \int_0^\infty \psi(S_T) f(S_T) \, dS_T,$$

where $\tau = T - t$ is the time to maturity, $r_{t,\tau}$ is the interest rate, $\psi(S_T) = \max\{S_T - X, 0\}$ is the payoff function, and $f$ is the state price density. For brevity, we leave implicit the dependence of $f$ on the horizon as well as other

Received May 2008; revised December 2008.
[1]Supported in part by NSF Grants DMS-0624841 and DMS-0706724.
*Key words and phrases.* Black–Scholes equation, European call options, nonparametric mixture, state price density.

economic variables such as the current asset price $S_t$, the interest rate and the dividend yield over the period.

The renowned Black–Scholes model assumes that the underlying asset price process $\{S_t\}$ follows a geometric Brownian motion and, therefore, the risk-neutral distribution is a log-normal distribution. Despite its elegance and popularity, it is now well understood that the log-normal assumption made by the Black–Scholes model can be problematic in practice and may result in severe bias of option prices. A number of econometric models have been developed to address this issue. Most notable examples include the stochastic volatility model and the GARCH model. The readers are referred to Garcia, Ghysels and Renault (2009) for a survey of recent developments in this direction. Although useful in a variety of contexts, these parametric models are still susceptible to model misspecification.

Various nonparametric methods have been employed to overcome this problem. Derman and Kani (1994), Dupire (1994) and Rubinstein (1994) propose implied binomial tree techniques to recover the state price density from a set of option prices without assuming the log-normality. Buchen and Kelly (1996) and Stutzer (1996) reconstruct the state price density under the maximum entropy principle. Jackwerth and Rubinstein (1996) introduce a smoothness penalized estimate. However, little is known about the econometric properties of these methods.

The state price density estimation is closely related to the recovery of the option pricing function $C$ itself. As observed by Banz and Miller (1978) and Breeden and Litzenberger (1978),

$$(1.2) \qquad f(S_T) = e^{r_{t,\tau}\tau} \frac{\partial^2 C}{\partial X^2}\bigg|_{X=S_T}.$$

Taking advantage of this relationship, the state price density can be derived as the second derivative of an estimate of the pricing function $C$. In the presence of pricing error, the estimation of the pricing function $C$ can be cast as a regression problem:

$$(1.3) \qquad C = C(X) + \varepsilon,$$

where, with slight abuse of notation, we use $C$ to denote the observed option price and $C(\cdot)$ to denote the correct pricing as a function of the strike price, and $\varepsilon$ represents the pricing error. Various nonparametric regression techniques have been applied to estimate $C(\cdot)$. In one of the pioneering papers, Hutchinson, Lo and Poggio (1994) consider estimating $C$ nonparametrically using various learning networks. More recently, Aït-Sahalia and Lo (1998) introduce a semiparametric alternative where the volatility of the Black–Scholes formulation is modeled nonparametrically. The readers are referred to Ghysels et al. (1997) and Fan (2005) for recent reviews of other nonparametric methods for estimating the option pricing function or the state price density.

From a statistical point of view, estimating the state price density now becomes estimating the second derivative of a regression function. But unlike other regression problems, the state price density needs to be a proper density function that is non-negative and integrates to the unity. This dictates, for example, that the price function $C(\cdot)$ is monotonically decreasing and convex in terms of the strike price $X$. More precisely,

$$-e^{r_{t,\tau}\tau} \leq C'(X) \leq 0,$$
$$C''(X) \geq 0.$$

How to impose these constraints presents the main difficulties in nonparametric regression (1.3). Aït-Sahalia and Duarte (2003) and Yatchew and Härdle (2005) stress the importance of enforcing such shape constraints in estimating the option pricing function and propose a nonparametric estimate of $C$ that respects these constraints. It is shown that both approaches lead to improved accuracy in recovering the pricing function and can guarantee the non-negativity of the state price density estimate. Neither state price density estimate, however, is guaranteed to integrate to one as required by a proper density. A post-estimate normalization is only necessary to ensure such constraint.

In this paper we develop a new approach to nonparametric estimation of the state price density. We propose to estimate the regression functions by minimizing the (weighted) least squares over a set of admissible pricing functions. The admissible pricing function is deducted directly from the very existence of a state price density. We consider a particular admissible set of pricing functions whose corresponding state price density is a nonparametric mixture of log-normals. We show that even though the minimization is taken over a infinite dimensional space, the minimizer actually admits a finite dimensional representation. In particular, all solutions can be expressed as a convex combination of at most $n+1$ Black–Schole type of pricing functions. In addition, we prove that, by focusing on the set of admissible pricing functions, not only the estimated state price density can be ensured to be a legitimate density function, but also the estimation accuracy can be drastically improved. More specifically, we show that as the sample size $n$ increases, the pricing function can be recovered with squared error converging to zero at the rate of $\ln^2 n/n$, which is very close to the $1/n$ convergence rate that is typically achieved only with much more restrictive parametric assumptions such as the log-normality. Further, we show that integrated squared error of the estimate of the state price density converges to zero at the rate of $\ln^4 n/n$, which again differs from the usual parametric rate only by a factor of the power of log sample size.

The rest of the paper is organized as follows. We describe the methodology in the next section. Section 3 discusses the asymptotic properties of the proposed estimate of both the call option prices and the state price density. The proposed estimating scheme is illustrated through an empirical study in Section 4. We close with some conclusions in the last section. All proofs are relegated to the Appendix.

**2. Method.** In the Black–Scholes paradigm, the state price density $f$ corresponds to a log-normal distribution. More precisely, the log return $\ln(S_T/S_t)$ follows a normal distribution with mean $(r_{t,\tau} - \delta_{t,\tau} - \sigma^2/2)\tau$ and variance $\sigma^2\tau$, where $\delta_{t,\tau}$ is the dividend yield in this period. Under this premise, (1.1) yields

$$(2.1) \qquad C(X, S_t, r_{t,\tau}, \tau) = S_t e^{-\delta_{t,\tau}\tau}\Phi(d_1) - X e^{-r_{t,\tau}\tau}\Phi(d_2),$$

where $\Phi(\cdot)$ is the cumulative distribution function of the standard normal distribution and

$$d_1 = \frac{\ln(S_t/X) + (r_{t,\tau} - \delta_{t,\tau} + \sigma^2/2)\tau}{\sigma\sqrt{\tau}},$$

$$d_2 = \frac{\ln(S_t/X) + (r_{t,\tau} - \delta_{t,\tau} - \sigma^2/2)\tau}{\sigma\sqrt{\tau}}.$$

The Black–Scholes formula (2.1) prices a European call option with only one parameter, $\sigma$, often referred to as the implied volatility. The Black–Schole model works remarkably well in the early years of option markets. It, however, becomes increasingly conspicuous that it fails to explain the option prices observed in the post-1987 crash market [Rubinstein (1994)]. To illustrate, in Figure 1, we plot a cross section of S&P 500 index option prices versus the strike price during a three week span in December 2002. The options expired on March 2003. We shall explain the main data characteristics in more detail in Section 4. Along with the observed data, we also plot the best fit given by the Black–Scholes model. It can be observed that the Black–Scholes model tends to underprice the deep in-the-money options. The discrepancy can be as much as 25% or about $20, which is rather significant.

Various approaches have been developed to improve the original Black–Scholes model. In the Black–Scholes paradigm, the underlying asset price is assumed to follow a geometric Brownian motion:

$$(2.2) \qquad dS = \eta S\, dt + \sigma S\, dB,$$

where $\eta$ is the growth rate of the stock price and $B_t$ is a standard Brownian motion. One popular alternative to the original Black–Scholes model is the

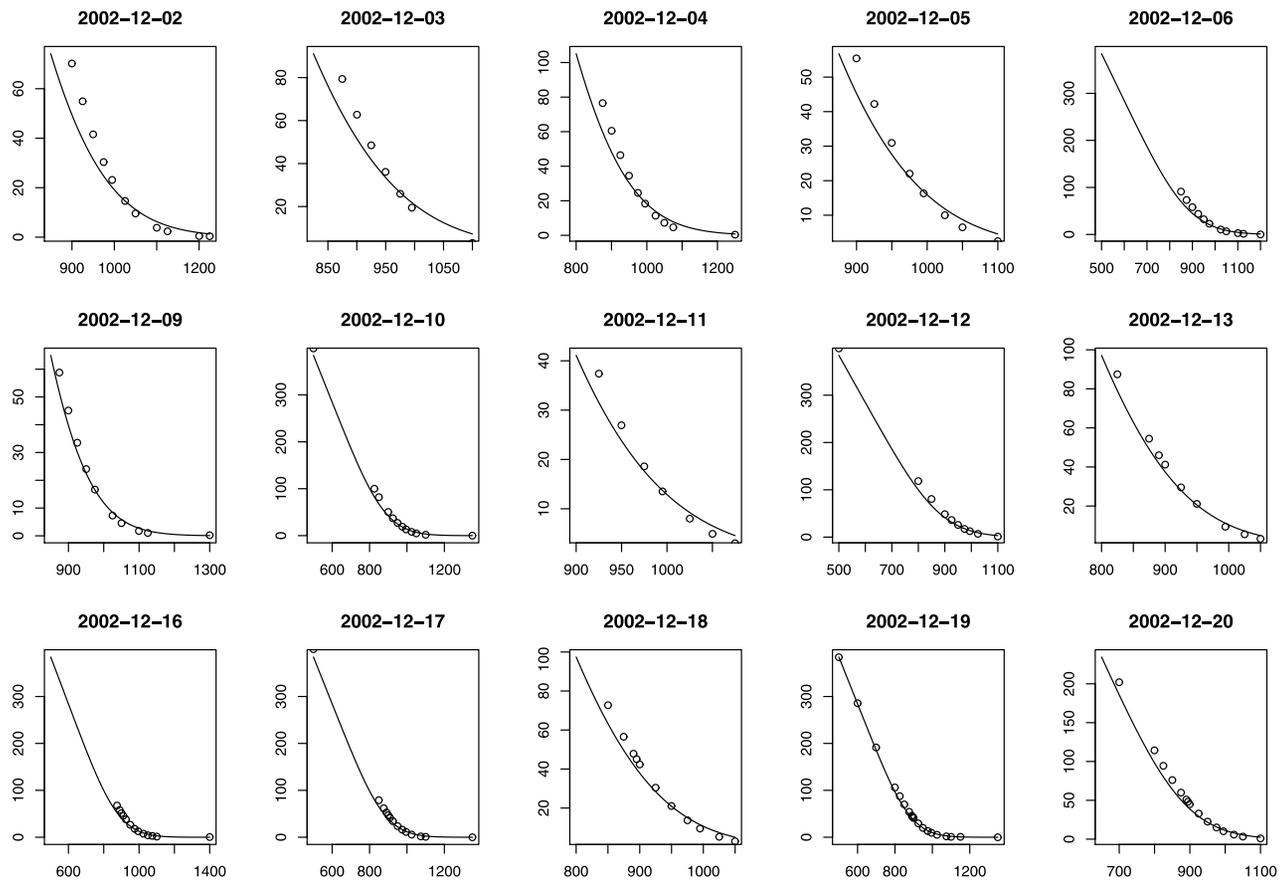

FIG. 1. *S&P 500 index option prices together with the best Black–Schole model fit.*

stochastic volatility model where $V = \sigma^2$ is further modeled as a stochastic process:

$$(2.3) \qquad dV = \zeta V \, dt + \xi V \, dW,$$

where $W$ is another standard Brownian motion. In the risk-neutral world, the asset price and its instantaneous variance $\sigma^2$ follow similar processes:

$$dS = (r - \delta) S \, dt + \sigma S \, d\tilde{B},$$
$$d\sigma^2 = \alpha \sigma^2 \, dt + \xi \sigma^2 \, d\tilde{W}.$$

Hull and White (1987) showed that if $\tilde{B}_t$ and $\tilde{W}_t$ are two independent Brownian motions, then the conditional distribution of $\ln(S_T/S_t)$ given "average" volatility

$$(2.4) \qquad \bar{V} = \frac{1}{T-t} \int_t^T \sigma^2(u) \, du$$

is normal with mean $(r_{t,\tau} - \delta_{t,\tau} - \bar{V}/2)\tau$ and variance $\bar{V}\tau$. More generally, using the same argument as Hull and White (1987), it can be shown that the statement holds true as long as $\tilde{B}_t$ is independent of the volatility process $V(t)$.

In other words, under the stochastic volatility model, $\ln(S_T/S_t)$ follows a mixture of normal distribution

$$(2.5) \quad f(\ln(S_T/S_t)) = \int \phi(\ln(S_T/S_t) | (r_{t,\tau} - \delta_{t,\tau} - \bar{V}/2)\tau, \bar{V}\tau) \, dF(\bar{V}),$$

where $\phi(\cdot | \mu, \sigma^2)$ is the normal density function with parameters $\mu$ and $\sigma^2$ and $F$ is the distribution function of $\bar{V}$. Clearly, this reduces to the Black–Scholes model when $F$ is a degenerated distribution. Motivated by this and to allow for more flexibility, we consider in this paper state price densities such that $\ln S_T$ follows a nonparametric mixture of normal densities:

$$(2.6) \qquad h(\ln S_T) = \int \phi(\ln S_T | \mu, \sigma^2) \, dG(\mu, \sigma),$$

where $G$, referred to as mixing distribution, is an unknown bivariate distribution function. The corresponding state price density can be written as

$$(2.7) \qquad f(S_T) = \int \upsilon(S_T | \mu, \sigma^2) \, dG(\mu, \sigma),$$

where $\upsilon(\cdot | \mu, \sigma^2)$ is the density function of log normal distribution with location parameter $\mu$ and scale parameter $\sigma$. It is evident that when $G$ assigns probability one to $(\ln(S_t) + (r_{t,\tau} - \delta_{t,\tau} - \sigma^2/2)\tau, \sigma\sqrt{\tau})$, the proposed mixture model reduces to the Black–Scholes model. The stochastic volatility model described above is also the special case of our nonparametric mixture model.

Different from the Black–Scholes models, our model for the state price density is nonparametric in that we do not impose any parametric assumption to the mixing distribution $G$. Mixtures of form (2.6) are known to be a rich family of distributions and can approximate any differentiable density function to an arbitrary precision [Silverman (1986)].

Our goal is to extract the state price density function $f$ as well as the pricing function $C$ given a set of observations on strike price and option price pairs, $(X_1, C_1), (X_2, C_2), \ldots, (X_n, C_n)$, that follow a regression relationship

$$(2.8) \qquad C_i = C(X_i) + \varepsilon_i, \qquad i = 1, 2, \ldots, n.$$

In particular, we assume that the state price density function lies in the following class:

$$(2.9) \qquad \mathcal{F} = \Big\{ f(\cdot) : f(S) = \int \upsilon(S|\mu, \sigma^2) \, dG(\mu, \sigma), \\ \operatorname{supp}(G) \subseteq [-M, M] \times [\underline{\sigma}, \bar{\sigma}] \Big\},$$

for some constants $M < \infty$ and $0 < \underline{\sigma} \leq \bar{\sigma} < \infty$. Correspondingly, the pricing function belongs to the following function class:

$$(2.10) \qquad \mathcal{C} = \Big\{ C(\cdot) : C(X) = e^{-r_{t,\tau}\tau} \int \psi(S) f(S) \, dS, f \in \mathcal{F} \Big\}.$$

Following Aït-Sahalia and Duarte (2003), we consider estimating the pricing function by minimizing the weighted least squares:

$$(2.11) \qquad \hat{C}(\cdot) = \arg \min_{C(\cdot) \in \mathcal{C}} \frac{1}{n} \sum_{i=1}^n w_i (C_i - C(X_i))^2.$$

As argued by Aït-Sahalia and Duarte (2003), the weights $w_i'$s can be chosen to reflect the relative liquidity of different options. More actively traded options would receive a higher weight than those less actively traded ones. They also suggest that the actual weights be determined on the basis of the size and time of the most recent transaction and the bid-ask spread, which are readily available in practice. For brevity, in the following discussion, we shall assume equal weights $w_1 = w_2 = \cdots = w_n = 1$. Our results, however, also apply to the more general and realistic weighting schemes.

Note that when the state price density is given by (2.7), the pricing function is also determined by the mixing distribution $G$:

$$C(X; G) = e^{-r_{t,\tau}\tau} \int \psi(S_T) f(S_T) \, dS_T \\ = e^{-r_{t,\tau}\tau} \int \psi(S_T) \int \upsilon(S_T|\mu, \sigma^2) \, dG(\mu, \sigma) \, dS_T$$

$$= e^{-r_{t,\tau}\tau} \int \left[ \int \psi(S_T) \upsilon(S_T|\mu,\sigma^2) \, dS_T \right] dG(\mu,\sigma)$$

$$\equiv \int C(X;\mu,\sigma^2) \, dG(\mu,\sigma),$$

where

$$\begin{aligned}
C(X;\mu,\sigma^2) &= e^{-r_{t,\tau}\tau} \int_0^\infty \psi(S_T) \upsilon(S_T|\mu,\sigma^2) \, dS_T \\
&= e^{-r_{t,\tau}\tau} \int_{-\infty}^\infty (e^s - X)_+ \phi(s|\mu,\sigma^2) \, ds \\
&= e^{-r_{t,\tau}\tau} \left( \int_{\ln X}^\infty e^s \phi_\sigma(s-\mu) \, ds - X \int_{\ln X}^\infty \phi_\sigma(s-\mu) \, ds \right) \\
&= e^{-r_{t,\tau}\tau + \sigma^2/2 + \mu} \left\{ 1 - \Phi\left(\frac{\ln X - (\mu + \sigma^2)}{\sigma}\right) \right\} \\
&\quad - e^{-r_{t,\tau}\tau} X \left\{ 1 - \Phi\left(\frac{\ln X - \mu}{\sigma}\right) \right\} \\
&= e^{-r_{t,\tau}\tau + \sigma^2/2 + \mu} \bar{\Phi}\left(\frac{\ln X - (\mu + \sigma^2)}{\sigma}\right) - e^{-r_{t,\tau}\tau} X \bar{\Phi}\left(\frac{\ln X - \mu}{\sigma}\right),
\end{aligned}$$

and $\bar{\Phi}(\cdot) = 1 - \Phi(\cdot)$.

The least squares estimate of the pricing function can be equivalently written as $\hat{C}(\cdot) = C(\cdot; \hat{G})$ with

$$(2.12) \qquad \hat{G}(\cdot) = \arg\min_{G \in \mathcal{G}} \frac{1}{n} \sum_{i=1}^n (C_i - C(X_i; G))^2,$$

where $\mathcal{G}$ is the collection of all probability measures on $\mu$ and $\sigma^2$. Note that the minimization is taken over a function space of infinite dimension, which is not directly computable at the first glance. However, as the following theorem shows, the solution can always be represented in a finite dimensional space and therefore make the minimization possible.

THEOREM 2.1. *The minimum of (2.12) exists and there is a distribution whose support contains no more than $n+1$ points achieves the minimum. Furthermore, at each support point, $\sigma^2 = \underline{\sigma}^2$.*

Theorem 2.1 is of great practical importance since it now suffices to find a minimizer of (2.12) that has a support of $n+1$ points or fewer, which can be solved numerically. The theorem is similar to theorems for optimal design [Silvey (1980)] and the the famous result for mixture likelihood [Lindsay (1983)].

In practice, it is common to ensure that the expected value of the price of the underlying security under the risk neutral measure is equal to the forward price of the underlying. This constraint can be easily incorporated in our framework. Note that when the state price density $f$ comes from $\mathcal{F}$, the expected value of $S_T$ can be conveniently expressed as

$$(2.13) \qquad E_f(S) = \int e^{\mu+\sigma^2/2} \, dG(\mu, \sigma^2).$$

Denote by $F_{t,T}$ the forward price of the underlying security. Enforcing the aforementioned constraint means that instead of $\mathcal{F}$, we restrict our attention to the following family of densities:

$$(2.14) \qquad \mathcal{F}^* = \left\{ f(\cdot) : f \in \mathcal{F}, \int e^{\mu+\sigma^2/2} \, dG(\mu, \sigma^2) = F_{t,T} \right\}.$$

Note that $\mathcal{F}^*$ is a convex subset of $\mathcal{F}$. Theorem 2.1 remains true. For the same reason, in the subsequent theoretical development, we shall neglect such constraint for brevity. But it is noteworthy that all our discussion also applies to the situation when this constraint is in place with little notational changes.

**3. Theoretical properties.** Before stating the main theoretical results, we first describe a set of conditions for the pricing errors. Assume that the pricing errors are independent and satisfy the following:

(a) $E(\varepsilon_i) = 0$, for $i = 1, \ldots, n$;
(b) for some $\beta > 0$, $\Gamma > 0$,

$$(3.1) \qquad \sup_n \max_{1 \le i \le n} E(\exp(\beta \varepsilon_i^2)) \le \Gamma \le +\infty.$$

Both conditions are rather mild. Condition (a) indicates that the observed price is unbiased, which provides the basis for estimating the pricing function. Condition (b) concerns how fast the tail of error distributions decays. Distributions that satisfy Condition (b) are often called sub-Gaussian. When the pricing errors follow normal distributions, this condition is satisfied. In the most realistic situations, the pricing error is bounded and this condition is also trivially satisfied.

We now consider the property of the estimated price function.

THEOREM 3.1. *Let $\hat{C}_n$ be the minimizer of (2.11) over $\mathcal{F}$. Then under Conditions* (a) *and* (b) *there exist constants $L_0, C_0 > 0$ such that, for any $n \ge n_0$ and $L \ge L_0$,*

$$(3.2) \qquad P\left( \frac{\sqrt{n}}{\ln n} \|\hat{C}_n - C\|_n > L \right) \le n^{-C_0 L^2},$$

*where*

(3.3) $$\|\hat{C}_n - C\|_n^2 = \frac{1}{n}\sum_{i=1}^n \{\hat{C}(X_i)_n - C(X_i)\}^2.$$

*Consequently,*

$$\|\hat{C}_n - C\|_n^2 = O_p\left(\frac{\ln^2 n}{n}\right).$$

The convergence rate obtained in Theorem 3.1 is to be compared with the usual parametric situation such as the Black–Scholes model. When assuming that the state price density follows a log-normal distribution, the pricing function can be given as (2.1). The volatility can also be estimated, for example, by means of the least squares. Such procedure would lead to the usual parametric convergence rate that $\|\hat{C}_n - C\|_n^2 = O_p(1/n)$. Note that, assuming the state price density resides in a much more general family $\mathcal{F}$, the convergence rate we obtained here only differs from the parametric rate by $\ln^2 n$.

Next, we study the properties of the estimated state price densities.

THEOREM 3.2. *Denote $\rho_X$ the sampling density of strike prices $X_1, \ldots, X_n$. Let $\Omega$ be an open set such that*

(3.4) $$\min_{x \in \Omega} \rho_X(x) \geq L_0 > 0$$

*for some constant $L_0$. Then under Conditions* (a) *and* (b)

(3.5) $$\int_\Omega (\hat{f}_n - f)^2 = O_p\left(\frac{\ln^4 n}{n}\right).$$

Similar to the price function, the state price density estimate converges at a "nearly" parametric rate, now with an extra term $\ln^4 n$. Compared with the price function, the rate is slightly slower. It is typical in nonparametric statistics that differentiation results in slower convergence rate. Different from the usual nonparametric setting, however, the convergence rate deteriorates only by $\ln^2 n$. In contrast, for both approaches from Aït-Sahalia and Duarte (2003) and Yatchew and Härdle (2005), the price function can be estimated at convergence rate $n^{-2q/(1+2q)}$ and state price density at $n^{-2(q-2)/(1+2q)}$, both in the integrated squared error sense when assuming that the price function is $q$ times differentiable.

In characterizing the performance of the state price density, confining to a set such as $\Omega$ is often necessary. The state price density is estimated based on observed pairs of strike and call price. Because we rarely observe strikes from regions where $\rho_X$ is close to zero, it is impossible to estimate it well in these regions without further restrictions. In practice, this is irrelevant because the strikes are most often evenly distributed in a compact region around the asset price. Putting it in our notation, this amounts to $\rho_X$ being a uniform distribution and set $\Omega$ of Theorem 3.2 can be taken as the whole support of the distribution.

## 4. Numerical studies.

4.1. *Implementation.* Theorem 2.1 shows that, without loss of generality, the estimated state price density admits the following expression:

$$\hat{f}(\ln(S_T/S_t)) = \pi_1 \phi(\ln(S_T/S_t); \mu_1, \underline{\sigma}^2) + \pi_2 \phi(\ln(S_T/S_t); \mu_2, \underline{\sigma}^2) \\ + \cdots + \pi_{n+1} \phi(\ln(S_T/S_t); \mu_{n+1}, \underline{\sigma}^2),$$

and it suffices to estimate the mixing proportions $\pi_j$s and means $\mu_j$s in minimizing the least squares. We propose to iteratively compute the mixing proportions and the means for a given $\underline{\sigma}^2$. Given $\mu_1, \ldots, \mu_{n+1}$, updating the mixing proportions can then be cast as a quadratic program and easily solved using the standard quadratic program solvers. Once the mixing proportions are available, we update the means by Newton Ralphson iterations.

The constraint that the expected value of the price of the underlying security under the state price density is equal to the forward price can also be easily incorporated in this algorithm, as it can now be conveniently expressed as

$$(4.1) \qquad F_{t,T} = S_t e^{\underline{\sigma}^2/2}(\pi_1 e^{\mu_1} + \cdots + \pi_{n+1} e^{\mu_{n+1}}).$$

It is of great importance to choose a good initial value. A careful examination of the proofs to Theorems 3.1 and 3.2 suggests that any density from $\mathcal{F}$ can be approximated well by a member of $\mathcal{F}$ but with the means $\mu$ to be equally spaced between $[-M, M]$. Motivated by this fact, we can take the means to be equally spaced as the initial value. The algorithm therefore starts with a natural initial solution, which is already a good estimate. A limited number of iterations are usually sufficient to achieve good performance in practical applications. We observe empirically that the least squares objective function decreases quickly in the first iteration, and the objective function after the first iteration is already very close to the objective function at convergence, as the magnitude of the decrease in the first iteration dominates the decreases in subsequent iterations. This motivates us to use a one-step iteration in our implementation.

4.2. *Simulation.* To gain insights to the finite sample performance of the proposed method, we first conducted a set of simulation studies. We adopted the experiment setting of Aït-Sahalia and Duarte (2003), which was designed to mimic S&P 500 index options. In particular, the current index price was set at 1365, the short term interest rate at 4.5%, the dividend yield at 2.5%, and the time to maturity at 30 days. The volatility smile was assumed to be a linear function of the strike with volatility equal to 40% at the strike price 1000 and 20% at the strike price 1700. We assume that we observe $n = 25$ option prices with strike prices equally spaced between 1000 and 1700. The option prices were simulated by adding uniformly distributed random noise to the theoretical option prices. Following Aït-Sahalia and Duarte (2003), the range of the noise varies linearly from 3% of the option value for deep in the money options to 18% for deep out of the money options.

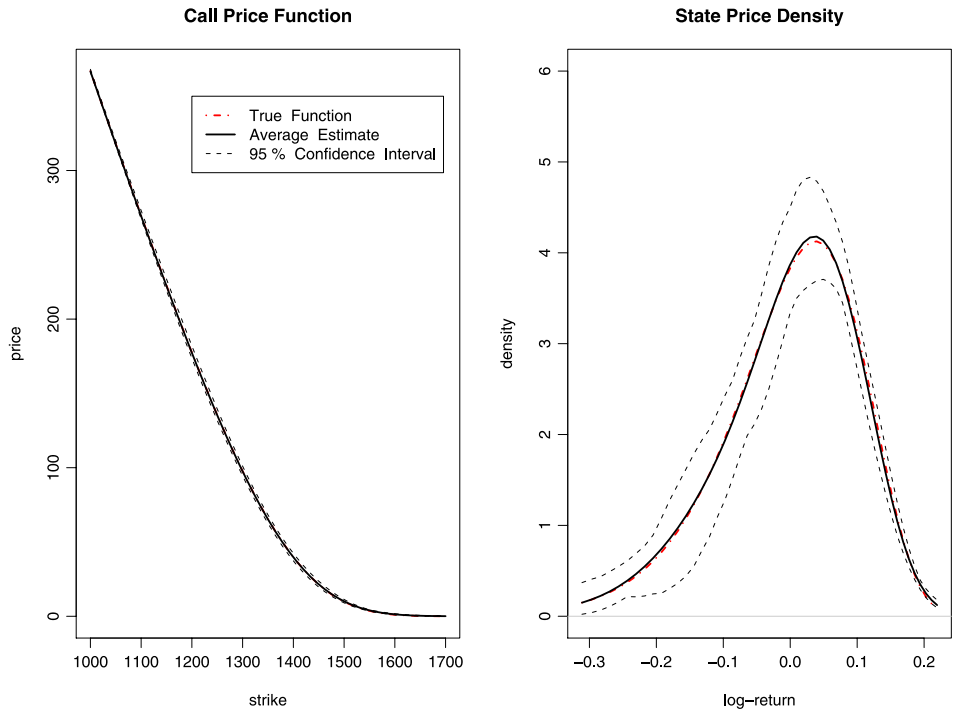

Fig. 2. *Estimated call option price function and state price density summarized over 5000 simulation runs.*

For each run, the call function and the corresponding state price density are estimated by the proposed method with $\underline{\sigma}$ chosen by leave one out cross validation [Wahba (1990)]. Figure 2 displays the average estimates and 95% pointwise confidence intervals for the call price function and the state price density based on 5000 simulations. It is evident that the estimate works very well.

4.3. *Real data analysis.* To illustrate the proposed methodology, we now go back to the historical option data briefly mentioned in Section 2. The data consist of a cross section of European call option prices written on the S&P 500 index during the first three weeks of December, 2002. Figure 3 shows the closing price of the index itself and the Eurodollar deposit rates (London) in the same period. The deposit rate is used as the risk-free interest rate. Because the maturity ranges from 3 months to about 4 months, we linearly interpolated the 3 month rate and 6 month rate to yield the daily risk-free rate.

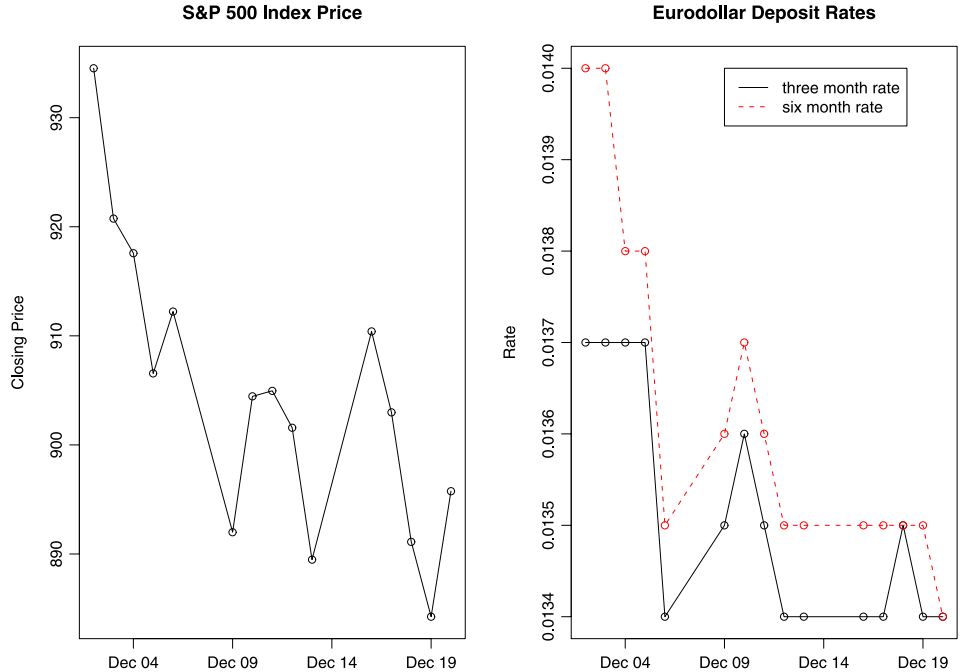

Fig. 3. *Historical S&P 500 index price and Eurodollar deposit rates from December 2, 2002 to December 20, 2002.*

The cross-section of the option prices are given in the leftmost panel of Figure 4. Following convention, we use the average of the end-of-day bid and ask price as the option price. Different lines correspond to different dates. It is clear that the option price can be modeled as a smooth function of the strike. This leads to the misperception that we can always estimate the state price density by directly differentiating an interpolation of the options prices. Such naïve strategy does not work in practice, however. To elaborate, the middle and right panels of Figure 4 show $\partial C/\partial X$ and $\partial^2 C/\partial X^2$ respectively estimated by straightforward differentiation. It can be observed that the derivatives are much more wiggly as functions of the strike and, furthermore, there is no guarantee that the resulting estimate of the state price density is positive as required by a legitimate density.

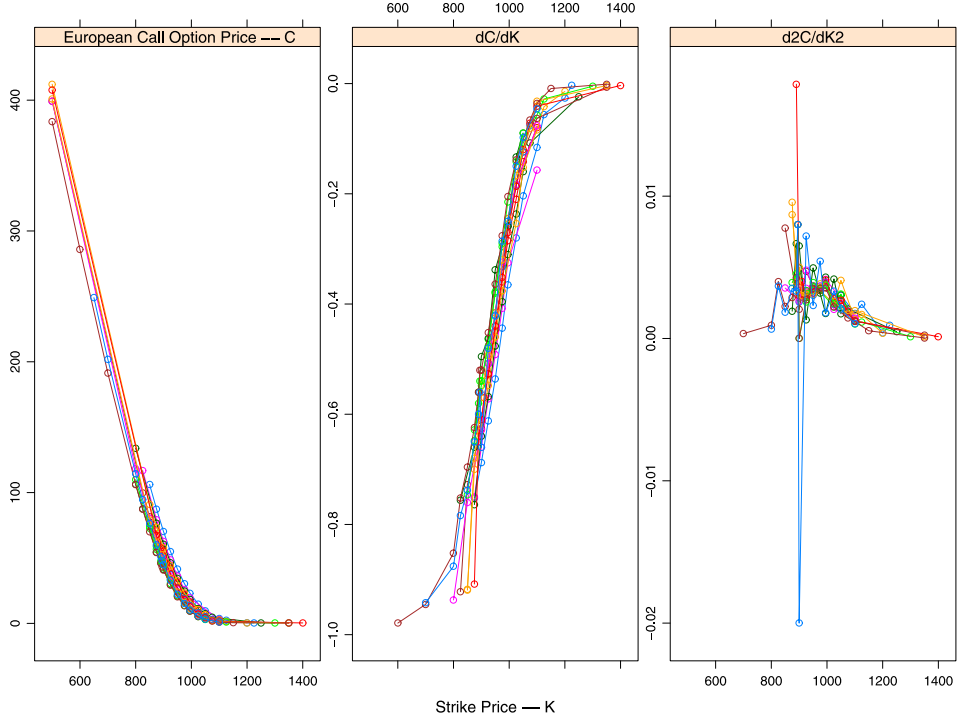

Fig. 4. *Historical European call option prices for S&P 500 index from December 2, 2002 to December 20, 2002 and its derivatives with respect to the strike. Different lines correspond to different dates.*

We now apply the proposed method to the option prices on a daily basis. As in Aït-Sahalia and Duarte (2003), we set the weights to be the inverse of the option price since the prices fluctuate considerably more when the price itself is high. The Black–Scholes model fit reported in Section 2 is produced in the same fashion. We also reconstruct the dividend rate through the put-call parity:

$$(4.2) \qquad P_t + S_t e^{-\delta \tau} = C_t + X e^{-r\tau}$$

using the put-call pair at the money. We choose $\underline{\sigma}$ using leave one out cross validation. Our experience suggests, however, usually three quarters of the volatility obtained from the Back–Scholes model fit works fairly well in practice. The estimated pricing functions and state price densities are given in Figures 5 and 6 respectively. In contrast to the Black–Schole model fit shown in Figure 1, our nonparametric estimate fits the historical option prices very well. The departure of the underlying state price densities from log normality is also evident from Figure 6.

These nonparametric state price density estimates can have many uses. For example, as pointed out in Aït-Sahalia and Duarte (2003), they can

STATE PRICE DENSITY ESTIMATION 15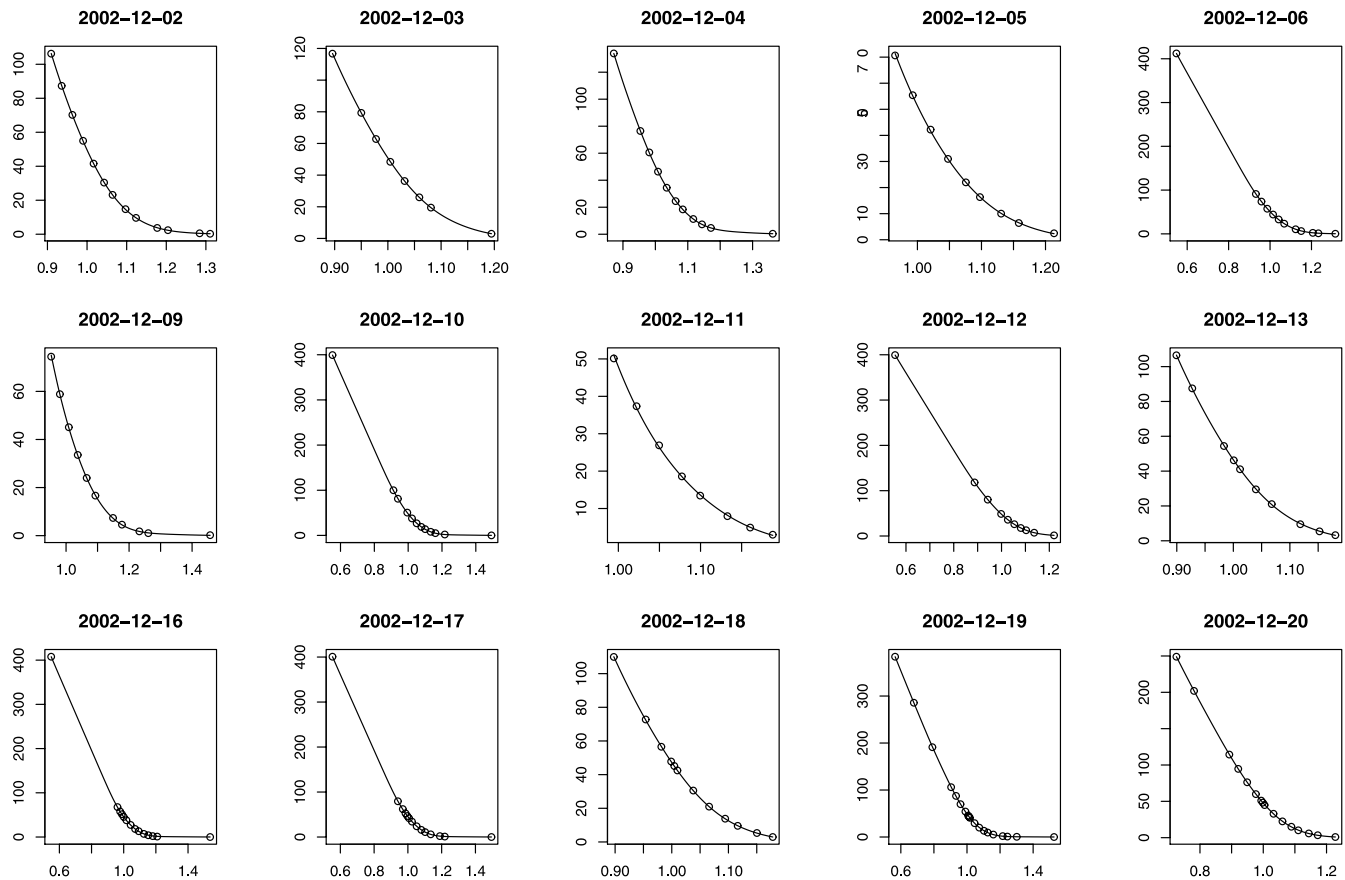

FIG. 5. *Estimated call prices versus Moneyness.*

STATE PRICE DENSITY ESTIMATION

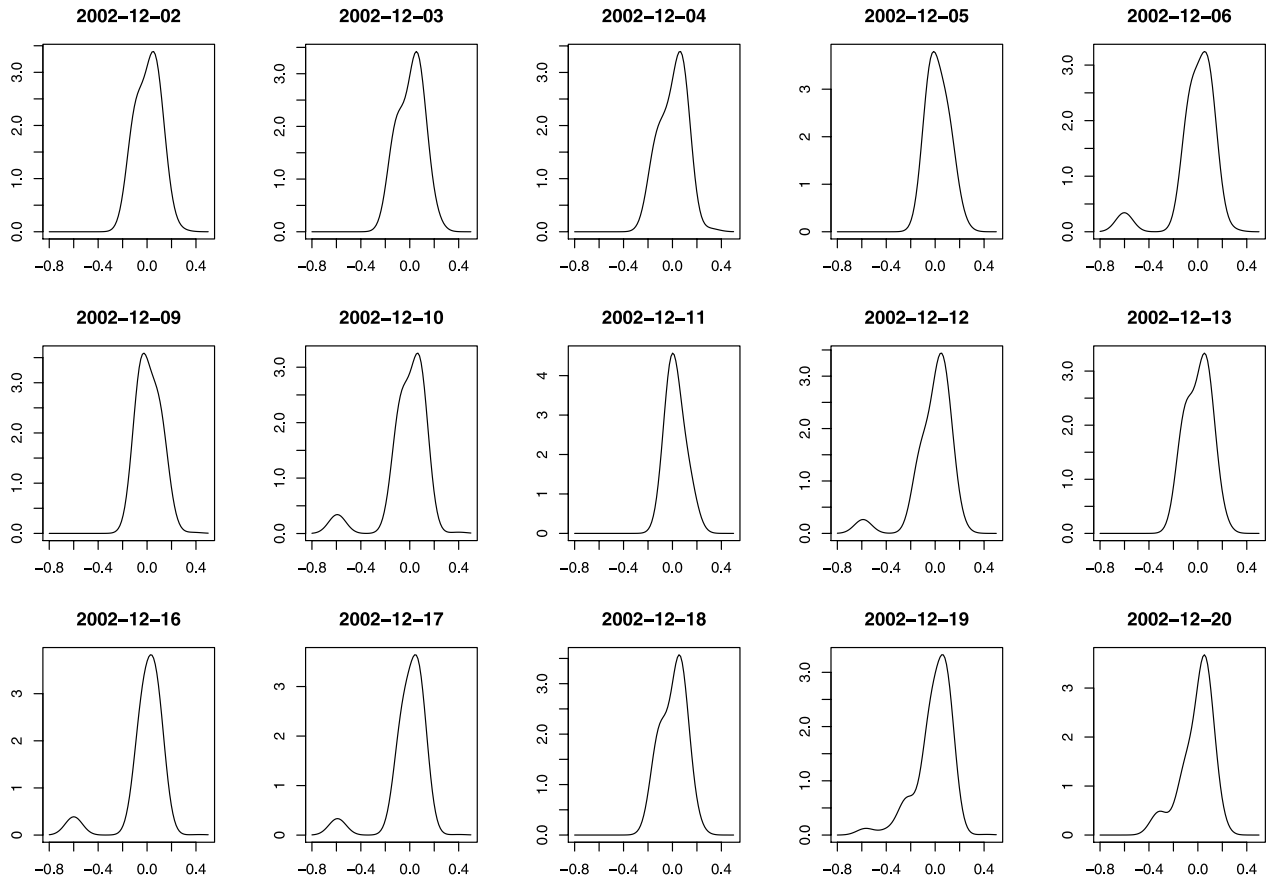

FIG. 6.   *Estimated state price density versus the excess log return* $\ln(S_T/S_t) - r_{t,\tau}$.

be employed to price new and more complex or less liquid options in an arbitrary-free fashion. With the knowledge of the entire state price density available, it is also straightforward to derive interesting quantities such as value-at-risk. Our nonparametric estimate can provide more reliable information than its parametric counterpart from these perspectives. For example, as evidenced in Figure 6, the Black–Scholes paradigm may significantly under-evaluate investment risk.

**5. Conclusions.** In this paper we introduced a new nonparametric option pricing technique. We consider state price densities that can be represented as a nonparametric mixture of log normals. Our nonparametric model is inspired by the stochastic volatility model of Hull and White (1987) and extends the original Black–Scholes model. Both the option price function and the state price density can be estimated through the least squares. We showed that such estimates enjoy nice asymptotic properties. An application to the historical data also demonstrates the merits of the proposed methodology in finite samples.

## APPENDIX

PROOF OF THEOREM 2.1. The existence of a minimum comes from the fact that both the objective function and the feasible region are convex. Denote

$$(5.1) \qquad \mathcal{A} = \{C(\cdot; \mu, \underline{\sigma}^2) : \mu \in \mathbf{R}\}.$$

It is not hard to see that $\mathcal{C}$ is a subset of the convex hull of $\mathcal{A}$. Let $\hat{G}$ be a minimizer of (2.12). Clearly, $(C(X_1, \hat{G}), C(X_2, \hat{G}), \ldots, C(X_n, \hat{G}))'$ is an element of the convex hull spanned by $\mathcal{A}$. By Carathéodory's theorem, there exits a subset $\mathcal{B}$ of $\mathcal{A}$ consisting of $n+1$ points or fewer so that $(C(X_1, \hat{G}), C(X_2, \hat{G}), \ldots, C(X_n, \hat{G}))'$ is in the convex hull spanned by $\mathcal{B}$. In other words, we can find $\mu_1, \mu_2, \ldots, \mu_{n+1} \in \mathbf{R}$ so that

$$(5.2) \qquad C(X_i, \hat{G}) = \sum_{j=1}^{n+1} \pi_j C(X_i; \mu_j, \underline{\sigma}^2) \qquad \forall i = 1, \ldots, n$$

for some $\pi_j \geq 0$ and $\pi_1 + \cdots + \pi_{n+1} = 1$. Therefore,

$$(5.3) \qquad G(\cdot) = \sum_{j=1}^{n+1} \pi_j \Phi(\cdot; \mu_j, \underline{\sigma}^2)$$

minimizes (2.12). □

PROOF OF THEOREM 3.1. Without loss of generality, we assume that $r_{t,\tau}$ is zero throughout the proof.

For $\varepsilon > 0$, the $\varepsilon$-covering number of $\mathcal{C}$, $\mathcal{N}(\varepsilon, \mathcal{C}, \|\cdot\|_\infty)$, is defined as the number of balls with radius $\varepsilon$ necessary to cover $\mathcal{C}$. Denote

$$\mathcal{H}(\varepsilon, \mathcal{C}, \|\cdot\|_\infty) = \ln \mathcal{N}(\varepsilon, \mathcal{C}, \|\cdot\|_\infty) \tag{5.4}$$

the $\varepsilon$-entropy of $\mathcal{C}$. In the light of Theorem 4.1 from van de Geer (1990), it suffices to show that, for any $\delta > 0$ small enough,

$$\int_0^\delta \mathcal{H}^{1/2}(u, \mathcal{C}, \|\cdot\|_\infty)\, du \leq C\delta \ln \frac{1}{\delta}. \tag{5.5}$$

We now set out to establish this inequality. We proceed by explicitly constructing an $\varepsilon$-covering set for $\mathcal{C}$.

An application of the Taylor expansion yields

$$\left| \phi(u) - \frac{1}{\sqrt{2\pi}} \sum_{j=0}^{k-1} \frac{(-1)^j u^{2j}}{2^j j!} \right| \leq \frac{1}{\sqrt{2\pi}} \frac{u^{2k}}{2^k k!} \leq \frac{1}{\sqrt{2\pi}} \left( \frac{eu^2}{2k} \right)^k. \tag{5.6}$$

For any $U > |u|$,

$$\bar{\Phi}(u) = \int_U^\infty \phi(z)\, dz + \int_u^U \phi(z)\, dz$$

$$= \int_U^\infty \phi(z)\, dz + \int_u^U \left( \frac{1}{\sqrt{2\pi}} \sum_{j=0}^{k-1} \frac{(-1)^j z^{2j}}{2^j j!} \right) dz$$

$$+ \int_u^U \left( \phi(u) - \frac{1}{\sqrt{2\pi}} \sum_{j=0}^{k-1} \frac{(-1)^j z^{2j}}{2^j j!} \right) dz.$$

Therefore,

$$\left| \bar{\Phi}(u) - \int_u^U \left( \frac{1}{\sqrt{2\pi}} \sum_{j=0}^{k-1} \frac{(-1)^j z^{2j}}{2^j j!} \right) dz \right| \leq \int_U^\infty \phi(z)\, dz + \int_u^U \frac{1}{\sqrt{2\pi}} \left( \frac{ez^2}{2k} \right)^k dz$$

$$\leq C\phi(U)/U + \frac{U^{2k+1} - u^{2k+1}}{\sqrt{2\pi}(2k+1)} \left( \frac{e}{2k} \right)^k$$

$$\leq \frac{C}{U} \left\{ \phi(U) + \left( \frac{eU^2}{2k} \right)^{k+1} \right\}.$$

Hereafter, we use $C > 0$ as a generic constant.

Now consider distribution functions $F$ and $G$ such that

$$\int u^j\, dF(u) = \int u^j\, dG(u), \qquad j = 1, \ldots, 2k-1. \tag{5.7}$$

Then,
$$\left|\int C(x;\mu,\sigma)\,dF(\mu) - \int C(x;\mu,\sigma)\,dG(\mu)\right|$$
$$= \left|\int C(x;\mu,\sigma)\,d\{F(\mu) - G(\mu)\}\right|$$
$$\leq \left|\int e^{\sigma^2/2+\mu}\bar{\Phi}\left(\frac{x-(\mu+\sigma^2)}{\sigma}\right) d\{F(\mu) - G(\mu)\}\right|$$
$$+ \left|e^x \int \bar{\Phi}\left(\frac{x-\mu}{\sigma}\right) d\{F(\mu) - G(\mu)\}\right|$$
$$\leq e^{\sigma^2/2+M}\left|\int \bar{\Phi}\left(\frac{x-(\mu+\sigma^2)}{\sigma}\right) d\{F(\mu) - G(\mu)\}\right|$$
$$+ e^x\left|\int \bar{\Phi}\left(\frac{x-\mu}{\sigma}\right) d\{F(\mu) - G(\mu)\}\right|$$
$$\leq e^{\sigma^2/2+M}\left|\int\left(\int_{(x-\mu-\sigma^2)/\sigma}^{U}\left(\frac{1}{\sqrt{2\pi}}\sum_{j=0}^{k-1}\frac{(-1)^j z^{2j}}{2^j j!}\right) dz\right) d\{F(\mu) - G(\mu)\}\right|$$
$$+ e^x\left|\int\left(\int_{(x-\mu)/\sigma}^{U}\left(\frac{1}{\sqrt{2\pi}}\sum_{j=0}^{k-1}\frac{(-1)^j z^{2j}}{2^j j!}\right) dz\right) d\{F(\mu) - G(\mu)\}\right|$$
$$+ (e^{\sigma^2/2+M} + e^x)\frac{C}{U}\left\{\phi(U) + \left(\frac{eU^2}{2k}\right)^{k+1}\right\}$$
$$= (e^{\sigma^2/2+M} + e^x)\frac{C}{U}\left\{\phi(U) + \left(\frac{eU^2}{2k}\right)^{k+1}\right\}.$$

Choosing $U = \sqrt{k/2}$ yields

(5.8) $$\left|\int C(x;\mu,\sigma)\,dF(\mu) - \int C(x;\mu,\sigma)\,dG(\mu)\right| \leq C(1+r)^{-k}/\sqrt{k}$$

for some $0 < r < \exp(1/2\sigma^2) - 1$ and all $x \leq \sqrt{k/2}$.

On the other hand, when $x > \sqrt{k/2}$,
$$\int C(x;\mu,\sigma)\,dF(\mu) \leq e^{\sigma^2/2+M}\int \bar{\Phi}\left(\frac{x-(\mu+\sigma^2)}{\sigma}\right) dF(\mu)$$
$$\leq e^{\sigma^2/2+M}\bar{\Phi}\left(\frac{x-M-\sigma^2}{\sigma}\right)$$
$$\leq Ce^{\sigma^2/2+M}\phi\left(\frac{x-M-\sigma^2}{\sigma}\right)/\left(\frac{x-M-\sigma^2}{\sigma}\right)$$
$$\leq Ce^{\sigma^2/2+M}e^{-k/2\sigma^2}/\sqrt{k}$$

$$\leq C(1+r)^{-k}/\sqrt{k}.$$

Subsequently, for any $x > \sqrt{k/2}$,

$$\left|\int C(x;\mu,\sigma)\,dF(\mu) - \int C(x;\mu,\sigma)\,dG(\mu)\right|$$

$$\leq \int C(x;\mu,\sigma)\,dF(\mu) + \int C(x;\mu,\sigma)\,dG(\mu)$$

$$\leq C(1+r)^{-k}/\sqrt{k}.$$

In summary, we conclude that

$$(5.9) \quad \left\|\int C(\cdot;\mu,\sigma)\,dF(\mu) - \int C(\cdot;\mu,\sigma)\,dG(\mu)\right\|_\infty \leq C(1+r)^{-k}/\sqrt{k}.$$

Lemma A.1 of Ghosal and van de Vaart (2001) shows that, for any distribution function $F$, there exists a probability measure $G$ supported on $2k$ points so that (5.7) is satisfied. In other words, we can always find a probability measure $G$ with support on at most $O(\ln(1/\varepsilon))$ points such that

$$(5.10) \quad \left\|\int C(\cdot;\mu,\sigma)\,dF(\mu) - \int C(\cdot;\mu,\sigma)\,dG(\mu)\right\|_\infty \leq \varepsilon/2.$$

Now note that, for $\varepsilon(>0)$ small enough,

$$|C(X;\mu,\sigma) - C(X;\mu+\varepsilon,\sigma)|$$

$$= \left|\int_{-\infty}^{+\infty}\{(e^s - X)_+ - (e^{s+\varepsilon} - X)_+\}\,d\Phi(s;\mu,\sigma^2)\right|$$

$$= \int_{\ln X}^{+\infty}(e^{s+\varepsilon} - e^s)\,d\Phi(s;\mu,\sigma^2) + \int_{\ln X - \varepsilon}^{\ln X}(e^{s+\varepsilon} - X)\,d\Phi(s;\mu,\sigma^2)$$

$$= (e^\varepsilon - 1)e^{\sigma^2/2+\mu}\left\{1 - \Phi\left(\frac{\ln X - (\mu+\sigma^2)}{\sigma}\right)\right\}$$

$$+ e^{\sigma^2/2+\mu+\varepsilon}\left\{\Phi\left(\frac{\ln X - (\mu+\sigma^2)}{\sigma}\right) - \Phi\left(\frac{\ln X - (\mu+\sigma^2+\varepsilon)}{\sigma}\right)\right\}$$

$$- X\left\{\Phi\left(\frac{\ln X - \mu}{\sigma}\right) - \Phi\left(\frac{\ln X - (\mu+\varepsilon)}{\sigma}\right)\right\}$$

$$\leq (e^\varepsilon - 1)e^{\sigma^2/2+\mu} + \varepsilon e^{\sigma^2/2+\mu+\varepsilon}/\sqrt{2\pi\sigma^2}$$

$$\leq \{(2 + 1/\underline{\sigma})e^{\bar{\sigma}^2/2+M}\}\varepsilon \equiv K\varepsilon.$$

In conclusion, for any $F \in \mathcal{F}$, we can always find a $G$ who is supported only on $0, \pm\varepsilon/K, \pm 2\varepsilon/K, \ldots, \pm[KM]\varepsilon/K$ such that

$$(5.11) \quad \left\|\int C(\cdot;\mu,\sigma)\,dF(\mu) - \int C(\cdot;\mu,\sigma)\,dG(\mu)\right\|_\infty \leq \varepsilon.$$

Together with the fact that there exists a generic constant $D > 0$ such that

(5.12) $$|C(X; \mu, \sigma) - C(X; \mu, \sigma + \varepsilon)| \leq D\varepsilon,$$

this implies the following bound on the covering number for $\mathcal{C}$:

(5.13) $$\mathcal{N}(\varepsilon, \mathcal{C}, \|\cdot\|_\infty) \leq C\left(\frac{1}{\varepsilon}\right)^{C \ln(1/\varepsilon)}.$$

Therefore, the entropy can be bounded as well:

(5.14) $$\mathcal{H}(\varepsilon, \mathcal{C}, \|\cdot\|_\infty) = \ln \mathcal{N}(\varepsilon, \mathcal{C}, \|\cdot\|_\infty) \leq C\left(\ln \frac{1}{\varepsilon}\right)^2.$$

Hence, for $\delta$ small enough,

(5.15) $$\int_0^\delta \mathcal{H}^{1/2}(u, \mathcal{C}, \|\cdot\|_\infty) \, du \leq C\delta \ln \frac{1}{\delta}.$$

The proof is now completed. □

PROOF OF THEOREM 3.2. Denote $h = \hat{C}_n - C$. From Theorem 3.1,

(5.16) $$\int_\Omega h^2 \leq \frac{1}{L_0} \int_\Omega h^2 \rho_X \leq \frac{1}{L_0} \int h^2 \rho_X = O_p\left(\frac{\ln^2 n}{n}\right).$$

It is not hard to see that $h'' = \hat{f}_n - f$. By Parseval's theorem,

(5.17) $$\int (h^{[k]})^2 = \int \|\mathcal{F}\{h^{[k]}\}\|^2,$$

where $\mathcal{F}\{h^{[k]}\}$ is the continuous Frourier transform of $h^{[k]}$. Furthermore, note that

(5.18) $$\mathcal{F}\{h^{[k]}\}(s) = s^{k-2} \mathcal{F}\{h''\}(s) = \sum_{j=1}^{n+1} \pi_j s^{k-2} \exp\left(-i\mu_j s - \frac{\sigma_j^2 s^2}{2}\right).$$

By the triangular inequality,

(5.19) $$\|\mathcal{F}\{h^{[k]}\}(s)\| \leq \sum_{j=1}^{n+1} \pi_j |s|^{k-2} \exp\left(-\frac{\sigma_j^2 s^2}{2}\right) \leq |s|^{k-2} \exp\left(-\frac{\underline{\sigma}^2 s^2}{2}\right).$$

Together with (5.17), we have

(5.20) $$\int (h^{[k]})^2 \leq \int s^{2(k-2)} \exp(-\underline{\sigma}^2 s^2) \, ds$$
$$= \sqrt{\frac{\pi}{\underline{\sigma}^2}} \frac{(2k-4)!}{2^{k-2}(k-2)!} (2\underline{\sigma}^2)^{-(k-2)}.$$

An application of the Kolmogorov interpolation inequality yields

$$\begin{aligned}
\int_\Omega (\hat{f}_n - f)^2 &= \int_\Omega (h'')^2 \\
&\leq C \left(\int_\Omega h^2\right)^{1-2/k} \left(\int_\Omega (h^{[k]})^2\right)^{2/k} \\
&\leq C \left(\int_\Omega h^2\right)^{1-2/k} \left(\int (h^{[k]})^2\right)^{2/k} \\
&= O_p\left(\left(\frac{\ln^2 n}{n}\right)^{1-2/k} \left(\frac{k}{2\underline{\sigma}^2}\right)^2\right).
\end{aligned}$$

The proof can now be completed by setting $k = \ln n$. $\square$

**Acknowledgments.** The author thanks the editor, the associate editor and two anonymous referees for comments that greatly improved the manuscript. This research was supported in part by grants from the National Science Foundation.

School of Industrial and Systems
Engineering
Georgia Institute of Technology
755 Ferst Drive NW
Atlanta, Georgia 30332
USA
E-mail: myuan@isye.gatech.edu